\documentclass[sigconf]{acmart}

\newcommand{\DatasetName}{IDImage}

\usepackage{colortbl}
\usepackage{bm}
\definecolor{ours}{RGB}{251, 242, 242}
\usepackage{multirow}
\usepackage{tabularx}
\usepackage{graphicx}
\usepackage{float}
\usepackage{setspace}
\usepackage{makecell}
\usepackage{hhline}

\definecolor{visual}{RGB}{241, 248, 242}
\definecolor{audio}{RGB}{244, 228, 223}
\definecolor{multi}{RGB}{251, 248, 233}
\definecolor{mlc_color}{RGB}{227, 233, 238}
\definecolor{bic_color}{RGB}{251, 242, 242}

\AtBeginDocument{%
  \providecommand\BibTeX{{%
    \normalfont B\kern-0.5em{\scshape i\kern-0.25em b}\kern-0.8em\TeX}}}

\renewcommand\footnotetextcopyrightpermission[1]{} %
\setcopyright{none}

\acmSubmissionID{0000}

\begin{document}

\title{Identity-Aware Vision-Language Model for Explainable Face Forgery Detection}

\author{Junhao Xu\textsuperscript{1}, Jingjing Chen\textsuperscript{1}, Yang Jiao\textsuperscript{1}, Jiacheng Zhang\textsuperscript{1}, Zhiyu Tan\textsuperscript{2}, Hao Li\textsuperscript{2}, Yu-Gang Jiang\textsuperscript{1}
}
\affiliation{%
  \institution{\textsuperscript{1}Shanghai Key Lab of Intell. Info. Processing, School of CS, Fudan University}
    \city{}
  \country{}}
\affiliation{%
  \institution{\textsuperscript{2}Fudan University}
    \city{}
  \country{}}
\email{junhaoxu23@m.fudan.edu.cn, chenjingjing@fudan.edu.cn,
yjiao23@m.fudan.edu.cn,}
\email{
jiachengzhang22@m.fudan.edu.cn,
zytan24@m.fudan.edu.cn,
lihao_lh@fudan.edu.cn,
ygj@fudan.edu.cn
}

\begin{abstract}
Recent advances in generative artificial intelligence have enabled the creation of highly realistic image forgeries, raising significant concerns about digital media authenticity.
While existing detection methods demonstrate promising results on benchmark datasets, they face critical limitations in real-world applications. First, existing detectors typically fail to detect semantic inconsistencies with the person’s
identity, such as implausible behaviors or incompatible environmental contexts in given images. Second, these methods rely heavily on low-level visual cues, making them effective for known forgeries but less reliable against new or unseen manipulation techniques.
To address these challenges, we present a novel personalized vision-language model (VLM) that integrates low-level visual artifact analysis and high-level semantic inconsistency detection. 
Unlike previous VLM-based methods, our approach avoids resource-intensive supervised fine-tuning that often struggles to preserve distinct identity characteristics. Instead, we employ a lightweight method that dynamically encodes identity-specific information into specialized identifier tokens. This design enables the model to learn distinct identity characteristics while maintaining robust generalization capabilities.
We further enhance detection capabilities through a lightweight detection adapter that extracts fine-grained information from shallow features of the vision encoder, preserving critical low-level evidence.
Comprehensive experiments demonstrate that our approach achieves 94.25\% accuracy and 94.08\% F1 score, outperforming both traditional forgery detectors and general VLMs while requiring only 10 extra tokens.
\end{abstract}


\maketitle

% Unnumbered footnote in the lower-left corner of the first page.
\begingroup
\renewcommand{\thefootnote}{}
\footnotetext{Code and data are available at: \url{https://github.com/xyyandxyy/IDImage}}
\addtocounter{footnote}{-1}
\endgroup

\section{INTRODUCTION}

\begin{figure}[t]
	\centering
	\includegraphics[width=0.45\textwidth]{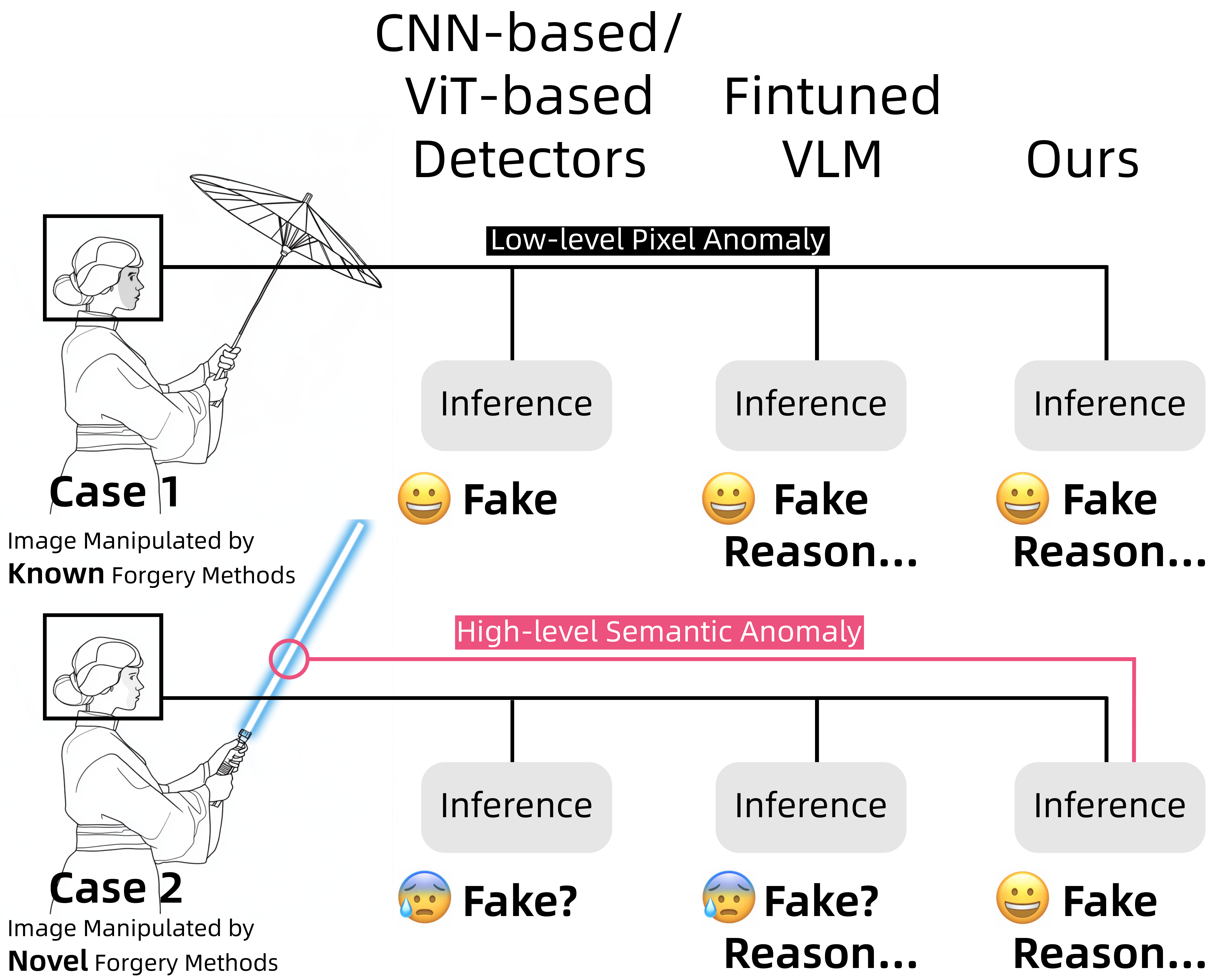}
\caption{While existing detectors and fine-tuned VLMs can handle known manipulation methods (Case 1), they usually struggle with novel forgery techniques (Case 2). Our approach leverages both low-level artifact detection and high-level semantic analysis through personalized identity priors, enabling robust detection with explanatory reasoning for both scenarios.}
        \label{teaser}
        \vspace{-5mm}
\end{figure} 

Recent advances in generative models, particularly GAN-based deepfakes and conditioned diffusion models, have enhanced image manipulation capabilities. They enable the effortless modification of identities, actions, and contextual elements with unprecedented realism and accessibility. These developments raise serious concerns about the authenticity of the media and the potential spread of visual misinformation.

In response to these emerging threats, the research community has developed various face forgery detection approaches, typically leveraging CNN-based and Vision Transformer architectures to identify manipulation artifacts. Despite promising results in benchmark datasets, current detection methods face three critical limitations that hinder their effectiveness in real-world applications.
First, existing approaches are unlikely to detect semantic inconsistencies related to identity, such as implausible behaviors, incompatible clothing, or incongruous environmental contexts for a specific person. As illustrated in Fig. \ref{teaser}, while conventional detectors may identify known manipulation artifacts (Case 1), they struggle with novel forgery methods (Case 2) where the manipulation appears visually plausible but contains semantic inconsistencies with the person's identity.
Second, existing detectors frequently struggle with generalization ability, exhibiting significant performance degradation when confronted with novel forgery techniques not represented in their training data. 
These limitations stem from an overreliance on low-level visual artifacts specific to particular generation algorithms while neglecting higher-level semantic inconsistencies that could serve as stronger detection cues.

Given these limitations, vision-language models (VLMs) offer promising capabilities for addressing the challenges in face forgery detection through their inherent high-level semantic modeling and cross-modal reasoning` abilities. 
However, naively applying supervised fine-tuning to VLMs for forgery detection presents significant challenges. 
First, while supervised fine-tuning offers a straightforward approach to enhance model performance in detecting image forgery, the simultaneous modeling of multiple individuals increases the challenge of preserving distinct identity characteristics, which leads to excessively additional annotated data for effective memorization and modeling of identity-specific representations.
Second, real-world deployment frequently necessitates continuous integration of new individuals, requiring persistent memorization of emerging identities. The conventional paradigm of recurrent model upgrade through direct fine-tuning not only imposes prohibitive temporal and computational burdens, but also introduces the risk of potential forgetting of previously learned representations.

To address these limitations, we introduce a novel personalized vision-language model built upon the LLaVA\cite{llava} framework to integrate both low-level artifact analysis and high-level personalized anomaly detection for comprehensive image forensics detection. 
Our approach requires only a small set of authentic reference images of an individual to embed forgery-sensitive prior knowledge into special identifier tokens and soft tokens.
We implement a lightweight strategy through personalized identifier tokens that effectively capture individual-specific visual characteristics, including appearance and behavior aspects. In contrast to supervised fine-tuning approaches, our method dynamically encodes identity-specific information. This strategy enables the model to learn distinct identity characteristics while maintaining generalization capabilities across diverse forgery detection scenarios.

The key insight driving our approach is that by combining injected personalized identity priors with VLMs' inherent commonsense reasoning capabilities, the model can effectively leverage high-level semantic cues for forgery detection. Our special identifiers serve dual roles: (1) enabling the VLM to identify identity-inconsistent information from input images and (2) guiding the generation of detailed, evidence-based forgery analysis explanations. To enhance the detection of low-level manipulation artifacts, we design a lightweight detection adapter that extracts fine-grained information from the shallow features of the VLM's vision encoder. This architectural enhancement enables more accurate and robust forgery detection across diverse manipulation techniques. It preserves critical low-level evidence typically dropped in deep feature hierarchies optimized for semantic understanding.

Our main contributions include:

\begin{itemize}
\item \textbf{Personalized Identity Priors}: We introduce an effective method for embedding personalized identity priors into pre-trained VLMs through specialized tokens, enabling the model to recognize identity-inconsistent elements without requiring extensive fine-tuning or large-scale annotation.
\item \textbf{Multi-level Feature Integration}: We propose a lightweight Detection Adapter that enhances the VLM's ability to capture low-level visual artifacts often overlooked by standard visual encoders while preserving the model's capacity to identify high-level semantic inconsistencies.
\item \textbf{Comprehensive Evaluation}: Extensive experiments demonstrate that our approach achieves 94.25\% accuracy and 94.08\% F1 score, outperforming both traditional forgery detectors and general VLMs by up to 5-11\% while requiring only 10 extra tokens. Our model maintains robust performance across both known and novel forgery techniques, providing detailed explanations that highlight specific inconsistencies rather than ambiguous visual artifacts.
\end{itemize}

\section{RELATED WORK}

\subsection{Face Forgery Detection}
Most Face forgery detection methods \cite{DBLP:conf/cvpr/LiL19c, DBLP:conf/icassp/YangLL19a, DBLP:conf/cvpr/LiuLZCH0ZY21, DBLP:conf/eccv/QianYSCS20, DBLP:conf/wifs/LiCL18, DBLP:conf/eccv/MasiKMGA20, DBLP:conf/ijcai/ZhangLL0G21, WildDeepfake, FTCN, DBLP:conf/icassp/ZhaoYNZ22, DBLP:conf/ijcai/HuXWLW021, DBLP:conf/aaai/GuCYDLM22} primarily focus on detecting visual defects and inconsistencies within facial regions to identify manipulated images. Beyond defect-based detection, researchers have explored identity-related cues for more robust forgery detection. For example, Dong \textit{et al.} \cite{OuterFace} introduced the use of identity-associated attributes as prior knowledge to learn discriminative identity embeddings. Extending this approach, Cozzolino \textit{et al.} \cite{ID-Reveal} developed an identity-aware framework that leverages a 3D morphable model capture facial representations, comparing suspect videos against reference videos using threshold-based discrimination. Further advancing this direction, Dong \textit{et al.} \cite{ICT} proposed an identity consistency transformer that effectively models identity representations across both inner and outer facial regions. Recently, Huang \textit{et al.} \cite{DBLP:conf/cvpr/HuangWYA00Y23} presented a framework that captures both explicit and implicit identity information through innovative contrastive and exploration losses.
Xu \textit{et al.} \cite{IDForge} proposed a reference-assisted framework that leverages identity information for multimedia forgery detection through dual contrastive learning strategies: identity-aware contrastive learning for identity-sensitive features and cross-modal contrastive learning to capture inconsistencies between modalities.

Although these methods achieve promising results on benchmark datasets, most of them function as black-box systems, lacking the transparent explanatory capabilities necessary for practical forensic applications. This limitation highlights the need for approaches that can not only detect identity inconsistencies but also provide interpretable explanations.

\subsection{Vision-Language Models for Detection}
Vision-Language Models (VLMs) integrate visual perception with language understanding \cite{gpt4o,llava,eagle,lumen,unitoken}, enabling cross-modal reasoning with applications across diverse domains. These models offer the potential for providing explanations, particularly when identifying semantic inconsistencies. While VLMs demonstrate impressive general knowledge capabilities—such as identifying well-known individuals—they typically lack specialized mechanisms for face forgery detection tasks. Jia \textit{et al.} \cite{Can_ChatGPT_Detect} evaluated GPT-4V and found it demonstrated moderate face forgery detection capability but exhibited notable challenges with real images and performed less effectively than existing detection methods. Niki \textit{et al.} \cite{foteinopoulou2024hitchhiker} conducted a systematic evaluation of four VLMs (BLIP-2, InstructBLIP, InstructBLIP-XXL, LLaVa-1.5) on face forgery detection tasks, revealing considerable limitations in generalization capacity.

To address these limitations, several specialized VLM-based approaches have been developed specifically for face forgery detection. Zhang \textit{et al.} \cite{DBLP:conf/eccv/ZhangCGSB24} utilized crowdsourcing platforms to collect human annotations for deepfake data and subsequently fine-tuned multimodal models such as BLIP to enhance detection performance. Similarly, recent frameworks, including FFAA \cite{ffaa} and FakeShield \cite{fakeshield}, have leveraged GPT-4o's capabilities to generate large-scale annotations and optimize model fine-tuning processes.

Despite these advancements, current VLM-based approaches for face forgery detection face several fundamental limitations. First, they primarily focus on detecting low-level visual artifacts through supervised fine-tuning, which deviates from the inherent strengths of VLMs in semantic reasoning. As demonstrated in Jia's research \cite{Can_ChatGPT_Detect}, even advanced models like GPT-4V struggle to identify subtle visual forgery artifacts effectively. Second, most current approaches rely on state-of-the-art VLMs like GPT-4o to generate training data descriptions based on minimal prompts and ground truth labels or forgery masks. Since these base VLMs themselves struggle with detecting visual artifacts, they often produce inaccurate descriptions that are treated as ground truth for fine-tuning, introducing false training signals.
Our approach addresses these limitations by fundamentally rethinking how VLMs should be leveraged for face forgery detection. Rather than forcing VLMs to identify low-level visual artifacts, we harness their natural strength in semantic understanding to detect identity inconsistencies. 
Inspired by research on personalizing VLMs for user-specific queries \cite{yollava, myvlm}, we encourage our VLM to use identity-related prior for forgery detection.
This strategy aligns with the identity-aware methods discussed earlier while offering two distinct advantages: (1) it preserves the models' original general reasoning capabilities, and (2) it enables the generation of interpretable, human-understandable explanations for detected forgeries.

\section{METHODOLOGY}

\subsection{Problem Formulation}

Given an input image $I$ of a specific identity, the goal is to determine whether it is authentic or manipulated and generate a textual explanation $T$ to justify the decision. We define our task as learning a function $f$ that maps:
\begin{equation}
f: (I, \{R_1, R_2, ..., R_n\}) \rightarrow (y, T),
\end{equation}
where $y \in \{0, 1\}$ indicates whether the image is real (0) or fake (1), $\{R_1, R_2, ..., R_n\}$ represents a small set of authentic reference images of the target identity, and $T$ is a detailed textual explanation outlining the evidence supporting the decision.

\begin{figure*}[ht]
	\centering
	\includegraphics[width=0.65\textwidth]{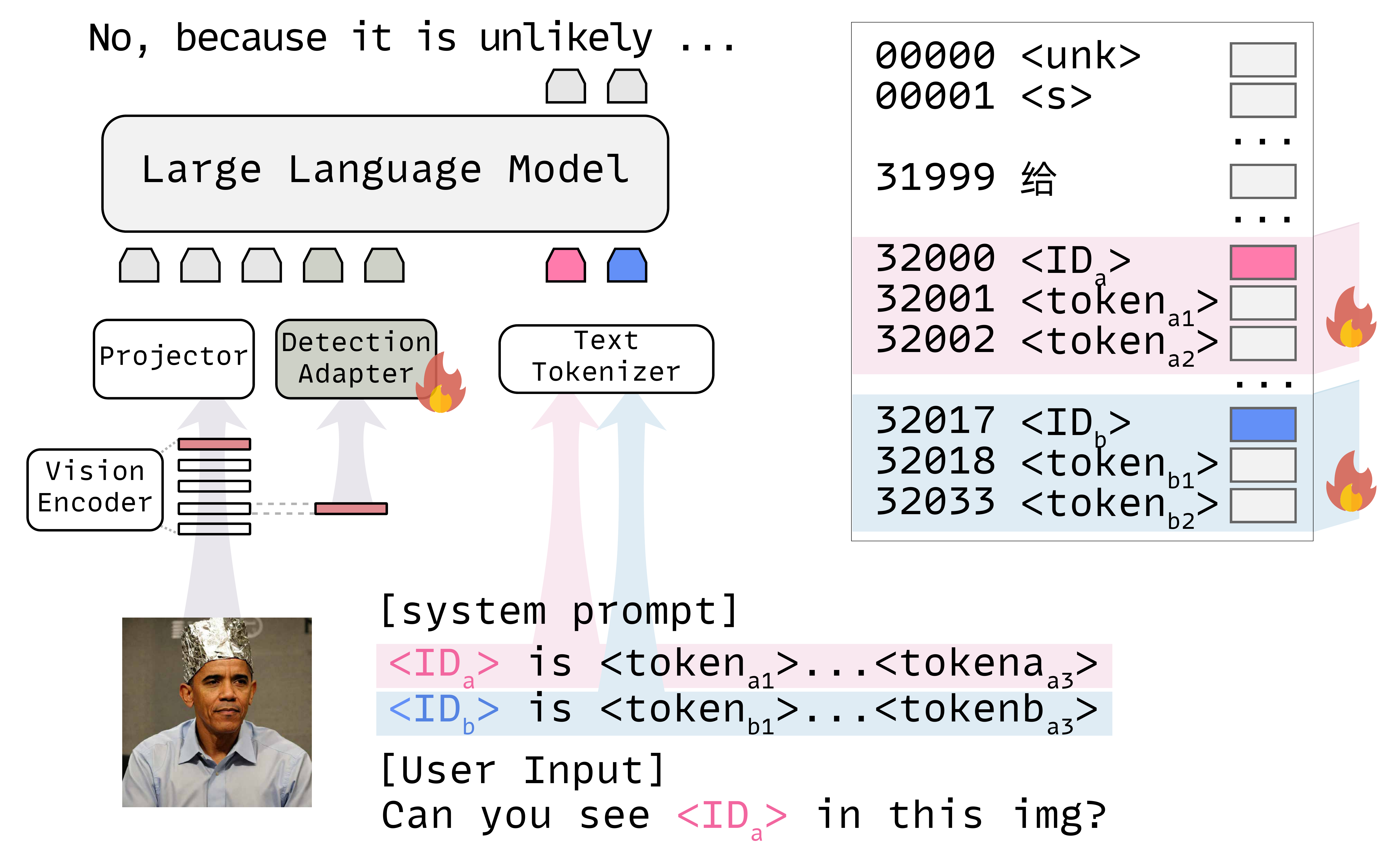}
	\caption{Overview of our proposed framework. Given a query image and a small set of authentic reference images, our approach (1) extracts personalized identity priors encoding both appearance and behavioral characteristics through specialized tokens (\texttt{<id\_a>} and \texttt{<id\_b>}), (2) leverages a lightweight Detection Adapter that preserves low-level visual artifacts from shallow layers of the vision encoder. The model can identify both visual inconsistencies (e.g., unnatural blending boundaries) and semantic implausibilities (e.g., inconsistent clothing or behavior) through this multi-level feature integration.}
        \label{model}
\end{figure*}

\subsection{Overview of Our Approach}

We propose a personalized vision-language model that integrates both low-level artifact analysis and high-level semantic inconsistency detection for comprehensive forgery detection. As illustrated in Fig.~\ref{model}, our framework consists of two key points:
(1) {Personalized Identity Prior Injection:} We encode identity-specific knowledge into specialized identifier tokens that enable the model to identify inconsistencies in appearance and behavior.
(2) {Detection Adapter:} A lightweight module that extracts fine-grained information from shallow layers of the vision encoder to preserve critical low-level evidence.
Our approach builds upon the LLaVA architecture, leveraging its vision-language capabilities while introducing forgery-sensitive components for the detection task.

\subsection{Personalized Identity Prior Injection}
Unlike previous VLM-based forgery detection methods that operate on generic visual patterns, our approach embeds personalized identity priors that enable the model to recognize identity-specific inconsistencies. The key insight is that by encoding these priors, the model can leverage high-level semantic understanding to identify manipulations.

\noindent\textbf{Representation of Identity Priors.} We represent identity priors using two specialized identifier tokens, denoted as \texttt{<id\_a>} and \texttt{<id\_b>}, which encode appearance prior and behavioral prior, respectively:

\begin{enumerate}
    \item \texttt{<id\_a>}: Encodes appearance-related attributes of the target identity, including facial features, hairstyle, clothing preferences, and other visual characteristics.
    \item \texttt{<id\_b>}: Captures behavioral patterns, typical contexts, and plausible activities associated with the target identity.
\end{enumerate}

\noindent As shown in Fig.\ref{model}, each identifier is assigned $N$ learnable soft tokens that encode a distributed representation of the identity's visual characteristics:
\begin{equation}
\begin{aligned}
\texttt{<id\_a>} &\rightarrow \{\texttt{<token\_a}_1\texttt{>}, \texttt{<token\_a}_2\texttt{>}, ..., \texttt{<token\_a}_N\texttt{>}\} \\
\texttt{<id\_b>} &\rightarrow \{\texttt{<token\_b}_1\texttt{>}, \texttt{<token\_b}_2\texttt{>}, ..., \texttt{<token\_b}_N\texttt{>}\}.
\end{aligned}
\end{equation}

\noindent This representation of identity priors offers several advantages over conventional text descriptions. First, soft tokens can capture subtle visual details that are difficult to express in language. Second, their distributed nature enables more precise discrimination between authentic and manipulated images of the target identity.

\noindent\textbf{Learning Identity Priors}.
To embed identity-specific knowledge into these tokens, we utilize a small set of authentic reference images $\{R_1, R_2, ..., R_n\}$ of the target identity. The training process involves two types of tasks:
(1) {Appearance Recognition:} Given an image, the model learns to determine whether it contains the target appearance by referencing the learned token representations.
(2) {Behavior Recognition:} The model learns to answer questions about behaviors of the target individual, encoding behavior-related information into the specialized tokens.

Formally, our trainable parameters include the specialized identifier tokens, their associated soft tokens, and the corresponding output weights in the language model's vocabulary matrix:

\begin{equation}
\theta_{\text{prior}} = \{\texttt{<id\_a>}, \texttt{<id\_b>}, \{\texttt{<token\_a/b}_i\texttt{>}\}_{i=1}^N, W_{\text{new}}\},
\end{equation}
where $W_{\text{new}}$ denotes the newly added rows in the output embedding matrix for the extended vocabulary tokens.

\subsection{Detection Adapter for Low-Level Artifact Preservation}

While specialized identifier tokens enable high-level semantic inconsistency detection, effectively identifying manipulation artifacts requires preserving low-level visual features that are often dropped in deeper layers of vision encoders. Rather than fine-tuning the entire vision encoder or introducing stronger vision models, we propose a lightweight Detection Adapter that extracts and preserves these critical low-level signals.

\noindent\textbf{Architecture Design}. Our Detection Adapter extracts features from a shallow layer of the vision encoder and projects them into visual tokens that can be directly processed by the language model:
\begin{equation}
T_{\text{adapter}} = \text{Proj}(F_{\text{shallow}}),
\end{equation}
where $F_{\text{shallow}}$ represents features from the initial layers of the vision encoder, and $\text{Proj}$ maps the features to token embeddings compatible with the language model. Such design offers several advantages:
(1) {Preservation of Low-Level Evidence:} shallow layers of vision transformers retain critical information about compression artifacts, noise patterns, and local inconsistencies that are typically strong indicators of manipulation.
(2) {Parameter Efficiency:} the adapter introduces minimal additional parameters compared to fine-tuning the entire vision encoder or introducing a separate model.
(3) {Preservation of Semantic Understanding:} by maintaining the original parameters of the VLM, we preserve its strong semantic understanding capabilities while enhancing its sensitivity to forgery artifacts.

The adapter tokens $T_{\text{adapter}}$ are combined with the standard visual tokens $T_{\text{standard}}$ (derived from the original vision-language projection) and fed into the language model:

\begin{equation}
T_{\text{integrated}} = [T_{\text{standard}}; T_{\text{adapter}}],
\end{equation}
where $[;]$ denotes concatenation along the sequence dimension. This allows the language model to leverage both high-level semantic information and low-level visual artifacts when generating forgery determinations and explanations.

\subsection{Training Strategy}

Training our model involves two primary stages: (1) learning the detection adapter and (2) personalizing the identity priors.

\noindent\textbf{Detection Adapter Training}.
To train the detection adapter, we collect a diverse dataset of authentic and manipulated images without requiring identity-specific information. This allows the model to learn generic forgery detection capabilities before personalization. We formulate this as a binary classification task with a simple Yes/No answer format to reduce training complexity:
\begin{equation}
\mathcal{L}\_{\text\{adapter\}} = \text{CrossEntropy}(\text{VLM}(T_{\text{standard}}; T_{\text{adapter}}), y).
\end{equation}
Where $T_{\text{adapter}}$ are the visual tokens produced by our adapter from shallow features, $T_{\text{standard}}$ are the standard visual tokens from the vision encoder, VLM produces a Yes/No response based on these visual tokens, and $y$ is the binary ground truth label. 
This binary response format simplifies the optimization landscape, stabilizing training while enabling efficient reuse of existing image forgery datasets. This stage leverages readily available data without requiring identity-specific information or additional annotation efforts, as binary labels can be obtained from existing metadata.

\noindent\textbf{Personalized Prior Training}.
For training personalized identity priors, we create a challenging dataset with authentic images of the target identity and high-quality forgery images. These forgeries are generated using state-of-the-art diffusion-based synthesis and GAN-based face-swapping models, producing realistic manipulations to help the model develop robust discrimination capabilities.
We design three types of training objectives:
\begin{enumerate}
    \item \textbf{Appearance Recognition:} a binary classification task determines if an image matches \texttt{<id\_a>} (appearance).
   \item \textbf{Behavior Recognition:} a binary classification task determines if an image is consistent with \texttt{<id\_b>} (behavior).
\end{enumerate}
The overall training objective is:
\begin{equation}
\mathcal{L}_{\text{total}} = \mathcal{L}_{\text{adapter}} + \lambda_1 \mathcal{L}_{\text{appearance}} + \lambda_2 \mathcal{L}_{\text{behavior}},
\end{equation}
where $\lambda_1$, $\lambda_2$, and $\lambda_3$ are hyperparameters controlling the relative importance of each objective.

\section{Dataset}
\subsection{Data Curation}
\noindent\textbf{Dataset Construction.} 
Due to the absence of identity-centric datasets specifically designed for VLM-based face forgery detection, we construct a novel dataset called  {\DatasetName} that incorporates both authentic and manipulated images of target individuals, along with corresponding authenticity labels and detailed descriptions for comprehensive evaluation.

{\DatasetName} dataset comprises images of 20 diverse individuals with varying ages and genders, selected from politics and entertainment domains. These English-speaking public figures are chosen due to their substantial media presence, making them statistically more vulnerable to digital impersonation attacks. For each individual, we curate approximately 279 real images, capturing them in diverse contexts, attire, and social settings, sometimes featuring multiple subjects within the same image to enhance complexity.

\noindent\textbf{Face Forgery Creations.} 
To generate face forgery creations, we employ five state-of-the-art manipulation techniques including both traditional deepfake methods and recent conditioned diffusion-based models. Using collected real images as reference, we generate face forgery creations that preserve the identity characteristics of the person while being entirely synthetic.

We divide the data into training and testing splits, as shown in Table. \ref{datasplit}.
For the training split, we utilized two representative image manipulation or generation methods: (1) SimSwap\cite{simswap}, and (2) PhotoMaker\cite{photomaker}. 
To better evaluate generalization capability, our test split exclusively contains forgeries created using entirely different methods: (3) Roop\cite{roop}, (4) StoryMaker\cite{storymaker}, and (5) PuLID\cite{pulid}. 
The dataset contains about 279 authentic and 780 manipulated images per individual in the training split, while the test split includes about 50 authentic and 50 manipulated images per individual.

\begin{table}
\centering
\caption{Dataset split with different image manipulation/generation methods}
\label{datasplit}
\begin{tabular}{clcl}
\hline
\textbf{Split} & \textbf{Methods} & \textbf{Type} & \textbf{Date}\\
\hline
\multirow{2}{*}{Training} & (1) SimSwap\cite{simswap} &Deepfake& 2021\\
                         & (2) PhotoMaker\cite{photomaker} &AIGC&2024\\
\hline
\multirow{3}{*}{Test}    & (3) Roop\cite{roop} & Deepfake&2022\\
                         & (4) StoryMaker\cite{storymaker} & AIGC&2024 \\
                         & (5) PuLID\cite{pulid} & AIGC&2024 \\
\hline
\end{tabular}
\end{table}

SimSwap and Roop are GAN-based face-swapping methods that transform a source face (a single reference image) into a target face while preserving the target's expressions and orientation.
StoryMaker and PuLID are conditioned diffusion-based methods that generate images guided by text prompts, using a set of reference images to maintain identity while creating new scenarios described in the prompts. The methods provided in the training set and test set are different. The deliberate distribution shift in {\DatasetName} mimics a challenging but realistic scenario, as detectors often perform poorly when training and testing on different data distributions.

\noindent\textbf{Description Annotation.} 
For images generated by diffusion models, we leverage the original generation prompts as a reference, as these prompts reflect the forgery creator's explicit intentions and manipulation directives. These prompts are then reformulated into standardized descriptive formats using an LLM (GPT-4o) to maintain consistency while preserving semantic content. For deepfake images, we employ a group of annotators to identify specific semantic anomalies across six dimensions: (1) hairstyle inconsistencies, (2) skin tone mismatches, (3) eyewear anomalies, (4) beard inconsistencies, (5) clothing style discrepancies, and (6) face shape distortions. Each image is evaluated from all six perspectives simultaneously in a multi-label approach, where multiple types of inconsistencies can be identified within the same image. 

\begin{figure}[t]
	\centering
	\includegraphics[width=0.5\textwidth]{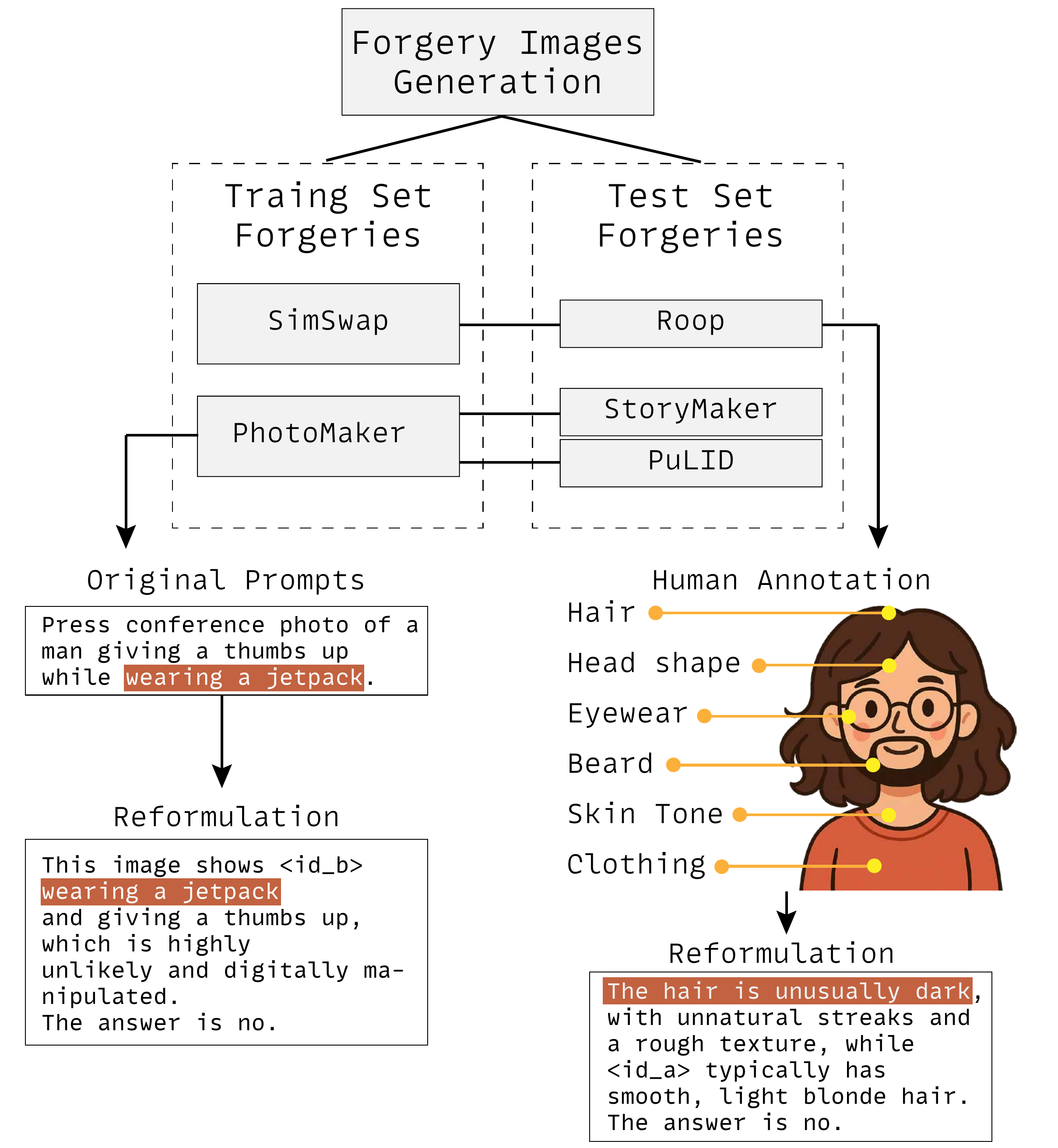}
	\caption{
Data Curation and Annotation.}
        \label{data_curation}
        \vspace{-5mm}
\end{figure}

\section{EXPERIMENTS}

\subsection{Experimental Setup}

\noindent\textbf{Implementation Details.}
In our implementation, we represent each individual with a total of 10 tokens, including identifiers <id\_a> and <id\_b>, with 4 soft tokens associated with each identifier. 
This optimal configuration is determined through experiments. 
We train the tokens for each individual separately.
We employ LLaVA-1.6-13B\cite{llava} as our base model, optimized using AdamW with a learning rate of 0.001. Each conversation is single-turn. 
The training dataset comprises approximately 279 real images per individual, along with 780 forgery images per individual (consisting of an average of 540 diffusion-generated images and 240 deepfake images). Although we use a substantial data volume per individual for a comprehensive evaluation, we later demonstrate that our approach remains effective with significantly reduced data quantities.
We train each subject for only 1 epoch in about 20 minutes. All experiments are conducted on NVIDIA A100 GPUs.

\noindent\textbf{Evaluation Metrics.}
For forgery detectors that produce binary results, we designate real images as positive samples and both AIGC and deepfake images as face forgery creations.
For all VLMs that produce natural language outputs, we use keyword retrieval to convert their responses into binary labels. Our evaluation metrics include image-level accuracy, precision, recall, and F1 score to provide a comprehensive assessment of classification performance.

\subsection{Baselines}

\noindent\textbf{Forgery Detectors.}
We compare our method with several representative forgery detection methods, including deepfake detectors (UCF\cite{ucf}, Recce\cite{recce}, Xception\cite{Xception}, and Meso4\cite{Meso}) and AI-generated content detectors (NPR\cite{NPR} and ESSP\cite{ESSP}). Xception and Meso are widely used deepfake detectors, while UCF and Recce are deepfake detection methods specially designed to enhance generalization performance. NPR and ESSP are recent methods that address not only GAN-generated deepfake images but also content generated by diffusion models. %

\noindent\textbf{Vision-Language Models.}
We also evaluate VLMs, including GPT4o-2024-08-06\cite{gpt4o}, Qwen2.5-VL\cite{qwen25vl}, and LLaVA-1.6-13B\cite{llava}. Following the settings in \cite{Can_ChatGPT_Detect}, we directly query these models with the prompt: "Tell me if there are synthesis artifacts in the face or not. Must return with yes or no only." Furthermore, we compared with FFAA\cite{ffaa}, a VLM specifically fine-tuned for face forgery detection, and Yo'LLaVA\cite{yollava}, a personalized VLM based on LLaVA. For FFAA and Yo'LLaVA, we employ the original prompts in their papers.

\subsection{Comparison with Baselines}
First, we compare the performance of our method with baselines on the collected test set. To ensure a fair comparison, we fine-tune the baseline forgery detectors using the collected training data. In addition, for VLM models, we also perform supervised fine-tuning on LLaVA using the entire training split. Note that in our test set, the forgery images are generated using methods different from those seen in the training data, presenting a challenging yet more realistic and practical setting.

The results are shown in Table \ref{sametraindata}. Among all traditional forgery detection methods (i.e., non-VLM-based approaches), ESSP achieves the best performance with a detection accuracy of 89.08\%, while Meso performs the worst with an accuracy of 55.17\%. 
This disparity can be attributed to the robust design of ESSP, which leverages noise-level analysis to effectively handle previously unseen data. 
On the other hand, Meso4, an earlier deepfake detection classifier, struggles with detecting fake patterns that were not part of its training dataset. 
Among all general VLMs, GPT-4o performs the best, achieving 83.03\% detection accuracy. However, LLaVA and Qwen2.5-VL perform substantially worse, with the latter showing perfect precision but extremely low recall (12.24\%), indicating that while it is accurate when detecting fake images, it seldom identifies them. FFAA, despite being specifically designed for face forgery detection, achieves only 41.67\% accuracy, highlighting the difficulty of generalizing to unseen forgery techniques.  
LLaVA with supervised fine-tuning achieves high recall (98.08\%) but moderate precision (60.71\%), suggesting it becomes overly sensitive to potential forgery indicators. Constrained by finite training samples, LLaVA appears to have developed a tendency to recognize patterns in response structures rather than learning generalizable features for forgery detection.
Yo'LLaVA shows high precision but only moderate recall (50.98\%).
This can be attributed to the fact that Yo'LLaVA was originally designed for general visual recognition tasks, causing it to face challenges when applied to the more complex domain of forgery detection.
Traditional detention methods generally perform better than VLM-based models. This performance difference can be attributed to the fact that traditional methods are specifically designed for the forgery detection domain. However, VLM-based models show more potential in usability as they can provide reasons for forgery judgment.

Our personalized approach effectively realizes this potential, combining the explanatory strengths of VLM-based models with significantly improved detection capabilities. Our method achieves an F1 score of 94.08\%. This indicates that our personalized approach can identify subtle inconsistencies even when faced with forgery techniques not seen during training.

\begin{table}[]
\caption{Performance comparision with baselines (\%).}%
\label{sametraindata}
\centering%
\small
\begin{tabular}{ccccc}%
\toprule%
Method&ACC&Precision&Recall&F1\\
\midrule%
Meso4 \cite{Meso}& 55.17& 53.38& 99.16 & 69.40\\
Xception \cite{Xception}&  80.03& 89.33& 69.31 & 78.06\\
Recce \cite{recce}&  75.09& 90.98& 57.06& 70.14\\
UCF \cite{ucf}&  81.69& 94.23& 68.48 & 79.32\\
NPR \cite{NPR} &  83.38& 92.96& 71.50 & 80.83\\
ESSP \cite{ESSP} &  89.08& 93.18& 83.86 & 88.27\\
\midrule%
Qwen2.5-VL \cite{qwen25vl} &  57.14& 100.0& 12.24& 21.81 \\
LLaVA \cite{llava} &  51.04& 50.68& 26.87 & 35.12\\
FFAA \cite{ffaa} &  41.67& 41.58& 47.64 & 44.36\\
Yo'LLaVA \cite{yollava}&  75.73 & 100.0 & 50.98 &67.53 \\
LLaVA (SFT) &  67.31& 60.71& 98.08 & 75.00\\
GPT4o \cite{gpt4o} &  83.03& 93.72& 70.07& 80.19 \\
\midrule%
Ours& 94.25 & 94.68& 94.12 & 94.08\\
\bottomrule%
\end{tabular}
\vspace{-5mm}
\end{table}

\subsection{Comparison with Personalized VLMs}
We then compare our method with VLMs by injecting personalization knowledge. To mimic a real-world scenario where a user describes a personalized subject to VLMs, we employ three methods to inject personalization knowledge for Qwen2.5-VL and GPT-4o: (1) Human-written description: human annotators manually write a description for each individual, simulating a real scenario where a user describes a personalized subject to VLMs. Each description is about 90 words ($\sim$ 150 tokens), including a rough description of appearance, clothing and styling, behavioral characteristics, and environmental context. (2) Reference Image: for each individual, we maintain a set of around 20 images featuring front-view photos without other people, clearly showing the identity. For each conversation, we randomly choose an image as a reference. (3) Reference Image + Human-written description: both human-written description and reference image are included in this situation.
For LLaVA, we use only the human-written description to inject personalization knowledge, as LLaVA can only process a single image as input.

\begin{table}[]
\caption{Comparison with Personalized VLMs (\%).}%
\centering%
\label{personalize_vlms}
\small
\begin{tabular}{cccccc}%
\toprule%
Method&ACC&Precision&Recall&F1&\makecell{Extra\\Tokens}\\
\midrule%
LLaVA & 51.04& 50.68& 26.87 & 35.12&0\\
LLaVA (Text) & 51.15& 50.08& 99.80 & 66.69& $\sim$ 150\\
\midrule%
Qwen2.5-VL & 7.14& 100.0& 12.24& 21.81 & 0\\
Qwen2.5-VL (Text) & 56.51& 52.89& 100.0 & 69.18&  $\sim$ 150\\
Qwen2.5-VL (Img) & 48.89& 48.89& 100.0 & 65.67& $\sim$ 10K\\
Qwen2.5-VL (All) & 50.53& 49.69& 100.0 & 66.39& $\sim$ 10K\\
\midrule%
GPT-4o & 83.03& 93.72& 70.07& 80.19 & 0\\
GPT-4o (Text) & 86.80& 77.22& 98.99 & 86.76& $\sim$ 150\\
GPT-4o (Img) & 65.77& 58.88& 99.90 & 74.09& $\sim$ 10K\\
GPT-4o (All) & 88.02& 80.95& 98.98 & 89.06& $\sim$ 10K\\
\midrule%
Ours&94.25 & 94.68& 94.12 & 94.08 & 10\\
\bottomrule%
\end{tabular}
\end{table}

The results are shown in Table \ref{personalize_vlms}. Generally, for all VLMs, injecting personalized knowledge helps improve the recall of the detection model. This is because introducing such information makes the model more sensitive to high-level inconsistencies, leading to an increase in recall. However, this improvement also results in a higher false positive rate, which causes precision to decrease compared to its baseline. Nevertheless, in most cases, the F1 score still shows substantial improvement. For LLaVA, the use of human-written descriptions raises the F1 score from 35.12\% to 66.69\%. For Qwen2.5-VL, the introduction of either human-written descriptions or reference images boosts the F1 score by more than 44\%. There, introducing additional tokens to encode personalization information proves to be beneficial for VLMs. Another notable observation is that, for both Qwen2.5-VL and GPT-4o, human-written descriptions lead to greater performance improvements than reference images. Specifically, Qwen2.5-VL achieves a detection accuracy of 56.15\% with human-written descriptions, compared to 48.89\% with reference images. Similarly, GPT-4o achieves 86.80\% accuracy when using human-written descriptions, while reference images result in a slight performance drop. These results suggest that single reference images provide limited personalization cues, as they capture only specific visual attributes from constrained angles and contexts. This limitation may cause models to overgeneralize from insufficient visual information. The combination of both human-written descriptions and reference images yields mixed results. For Qwen2.5-VL, this combination results in an F1 of 66.39\%, which is slightly lower than using human-written descriptions alone. This suggests that the model may struggle to effectively integrate visual and textual information. In contrast, GPT-4o achieves its best performance with this combination, reaching an F1 score of 89.06\%. This indicates a stronger capability in aligning and leveraging multimodal information compared to other models. Our method achieves the highest F1 score of 94.08\% while using only 10 extra tokens—significantly fewer than all other approaches. This demonstrates the effectiveness of our approach in capturing and utilizing identity information for personalized detection.

\begin{figure*}[t]
	\centering
	\includegraphics[width=1\textwidth]{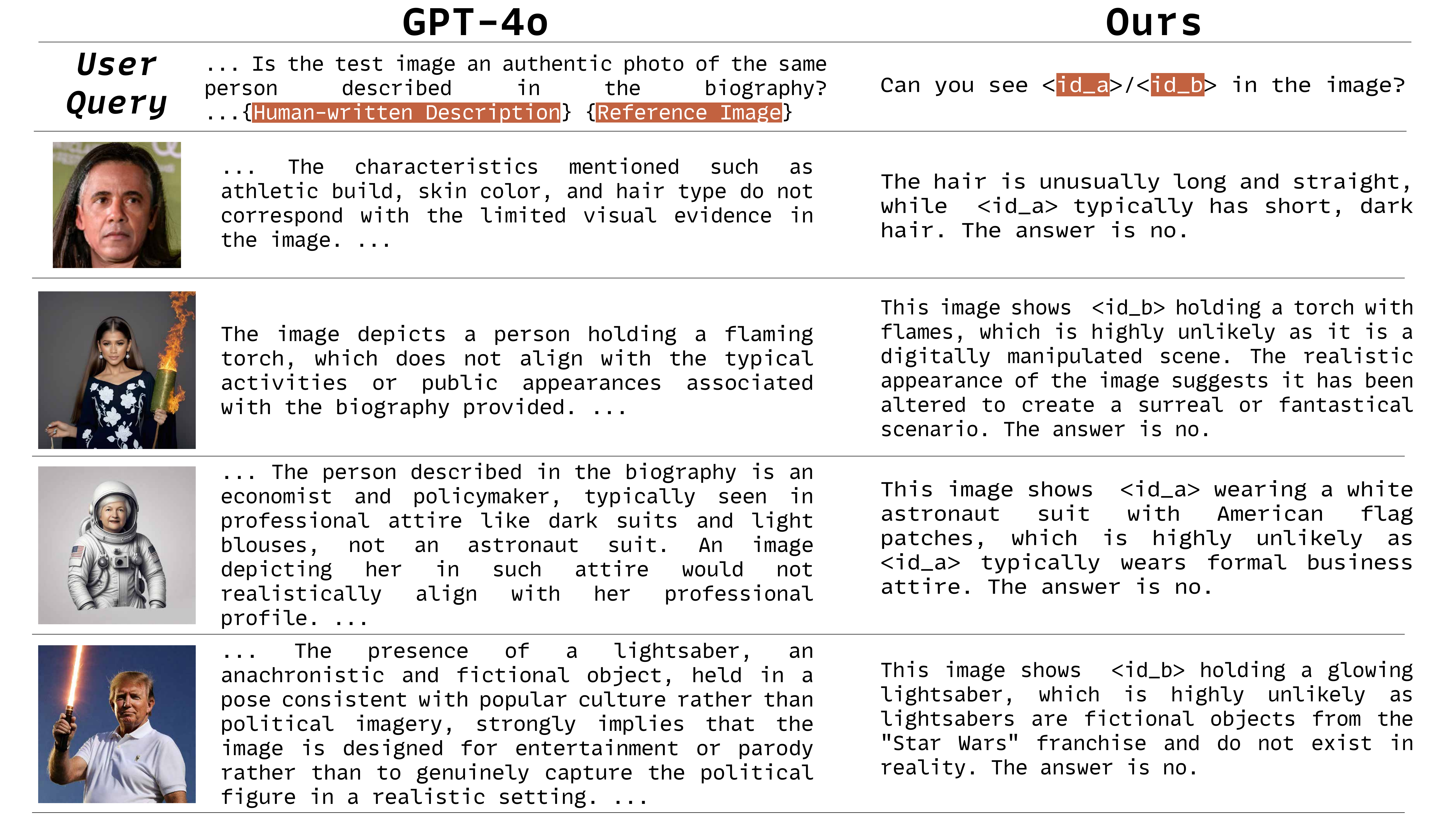}
    
	\caption{Qualitative comparison of forgery detection interpretability between our method and GPT-4o}
        \label{showcase}
\end{figure*}
\subsection{Ablation Study}
\noindent\textbf{Effect of Detection Adapter.}
To evaluate the contribution of our Detection Adapter module, we compare our full model against a variant where the Detection Adapter is removed. In this ablation, input images are directly processed by the ViT encoder, producing hidden states that are transformed into visual tokens by only the original adapter. The Personalized Identity Prior Injection remains operational in both configurations.
As shown in Table \ref{ablation}, removing the Detection Adapter (w/o Detection Adapter) increases recall but decreases other metrics, particularly Precision. This suggests the model loses important low-level visual cues and becomes biased toward classifying more inputs as forgeries, reducing its overall discriminative ability.

\noindent\textbf{Effect of Decoupled Specialized Identifier Tokens.}
We investigate whether using separate tokens for appearance and behavior priors (<id\_a> and <id\_b>) provides advantages over a single unified identity token (<id>). As shown in Table \ref{ablation}, our results (w/o Decoupled Tokens) demonstrate that using decoupled tokens yields better performance compared to the unified approach. 

\noindent\textbf{Effect of Chain-of-thought Reasoning.}
We evaluated a variant that produces only binary decisions without chain-of-thought reasoning. As shown in Table \ref{ablation}, removing the description (w/o CoT) in training performs worse, indicating that training the model with chain-of-thought helps learn more effective features. The reasoning process may serve as a condition for the binary decision, potentially helping the model consider relevant evidence before classification.

\begin{table}[]
\caption{Ablation Study (\%).}%
\label{ablation}
\centering%
\small
\begin{tabular}{ccccc}%
\toprule%
Method&ACC&Precision&Recall&F1\\
\midrule%
Ours(w/o Adapter)&82.42&76.56&98.00&85.96\\
Ours(w/o Decoupled Tokens)&87.09&88.27&87.62&86.33\\
Ours(w/o CoT)&89.09&93.51&84.03&87.26\\
Ours&94.25 & 94.68& 94.12 & 94.08\\

\bottomrule%
\end{tabular}
\end{table}

\subsection{Discussion on the Training Data Size} 
In our implementation, we use approximately 1,000 training samples per individual (279 positive samples and 780 face forgery creations) to fully explore our VLM's capabilities. However, in practical applications, collecting such extensive data for each individual is costly and labor-intensive, particularly gathering 200-300 real images per person. Therefore, we conduct experiments to evaluate our approach's performance across different data scales.

Table \ref{datascale} presents the performance of our model across different training data proportions. The results show that performance improves consistently as the amount of training data increases. With only 5\% of the data, the model achieves an F1 score of 68.00\%, which improves substantially to 81.73\% when using 10\% of the data. At 25\% data utilization, the model reaches an F1 score of 86.94\%, representing an improvement that may be sufficient for many practical applications. As we continue to increase the data proportion, performance continues to improve, with the F1 score reaching 86.69\% at 50\% data, 90.66\% at 75\%, and 94.08\% with the complete dataset.

\begin{table}[]
\caption{Impact of Training Data Size}%
\label{datascale}
\centering%
\small
\begin{tabular}{ccccc}%
\toprule%
Training Data&ACC&Precision&Recall&F1\\
\midrule%
5\%& 69.21& 72.34& 72.54 & 68.00\\
10\%& 80.43& 80.97& 85.41 & 81.73\\
25\%& 85.45& 83.00& 92.85 & 86.94\\
50\%& 86.80& 86.95& 89.17 & 86.69\\
75\%& 89.93& 89.30& 93.30 & 90.66\\
100\%&94.25 & 94.68& 94.12 & 94.08\\
\bottomrule%
\end{tabular}
\end{table}

\subsection{Evaluation Explanation Quality}
To evaluate the quality of our model's explanations, we perform human evaluation on the model's output following the protocol in \cite{foteinopoulou2024hitchhiker}. In the evaluation, 20 participants are asked to rate model responses on a 5-point scale based on their relevance to key issues. Clarity, conciseness, and communicative effectiveness are also emphasized in the evaluation. During the evaluation, 400 samples are randomly selected and fed into the model to obtain responses. Each participant evaluates 20 samples. Table \ref{explanation} summarizes the results.

In general, our model achieves the highest rating score of 4.01, significantly outperforming GPT-4o and Qwen 2.5-VL. This demonstrates its ability to generate concise and evidence-based explanations, as illustrated in Figure \ref{showcase}. GPT-4o achieves a score of 3.87, slightly lower than ours. While GPT-4o is capable of producing relevant and informative explanations, its outputs are often overly verbose, which affects clarity and conciseness in human evaluation. 
In contrast, Qwen 2.5-VL receives the lowest score of 2.91. The lower performance can be attributed to its tendency to produce generic or overly brief responses, often lacking concrete reasoning tied to the reference identity or specific forgery evidence, thus limiting its interpretability and practical value.

\begin{table}[]
\caption{Evaluation of Model Explanation Quality.}%
\centering%
\label{explanation}
\small
\begin{tabular}{ccc}%
\toprule%
Method &Human Evaluation \\
\midrule%
Qwen2.5-VL (All) &  2.91/5.00 \\
GPT-4o (All) & 3.87/5.00\\
\midrule%
Ours&4.01/5.00&  \\
\bottomrule%
\end{tabular}
\vspace{-3mm}
\end{table}

\section{CONCLUSION}
In this paper, we present a novel personalized vision-language model for robust face forgery detection. 
Our approach addresses critical limitations in current methods by integrating low-level artifact analysis with high-level semantic reasoning. 
By integrating specialized identifier tokens that efficiently encode identity-specific information using only a small set of reference images, our approach enables effective detection of semantic inconsistencies without extensive fine-tuning requirements. The lightweight detection adapter extracts crucial low-level evidence typically dropped in deep feature hierarchies while preserving the VLM's original semantic understanding capabilities.
Experiments on our {\DatasetName} dataset demonstrate that our method significantly outperforms both traditional forgery detectors and VLMs while requiring minimal additional parameters. Our approach maintains robust performance even with substantially reduced training data, demonstrating strong results with only a fraction of the full dataset. 
Our method sheds light on real-world applications requiring personalized and explainable forgery detection.

\bibliographystyle{ACM-Reference-Format}
\bibliography{reference}

@misc{roop,
  author = {s0md3v},
  title = {Roop},
  year = {2023},
  publisher = {GitHub},
  howpublished = {\url{https://github.com/s0md3v/roop}}
}

@inproceedings{simswap,
  author    = {Renwang Chen and
               Xuanhong Chen and
               Bingbing Ni and
               Yanhao Ge},
  title     = {SimSwap: An Efficient Framework For High Fidelity Face Swapping},
  booktitle = {{MM} '20: The 28th {ACM} International Conference on Multimedia},
  year      = {2020}
}

@inproceedings{DBLP:conf/icassp/YangLL19a,
  author    = {Xin Yang and
               Yuezun Li and
               Siwei Lyu},
  title     = {Exposing Deep Fakes Using Inconsistent Head Poses},
  booktitle = {{IEEE} International Conference on Acoustics, Speech and Signal Processing,
               {ICASSP} 2019, Brighton, United Kingdom, May 12-17, 2019},
  pages     = {8261--8265},
  publisher = {{IEEE}},
  year      = {2019},
  url       = {https://doi.org/10.1109/ICASSP.2019.8683164},
  doi       = {10.1109/ICASSP.2019.8683164},
  bibsource = {dblp computer science bibliography, https://dblp.org}
}

@inproceedings{DBLP:conf/wifs/LiCL18,
  author    = {Yuezun Li and
               Ming{-}Ching Chang and
               Siwei Lyu},
  title     = {In Ictu Oculi: Exposing {AI} Created Fake Videos by Detecting Eye
               Blinking},
  booktitle = {2018 {IEEE} International Workshop on Information Forensics and Security,
               {WIFS} 2018, Hong Kong, China, December 11-13, 2018},
  pages     = {1--7},
  publisher = {{IEEE}},
  year      = {2018},
  url       = {https://doi.org/10.1109/WIFS.2018.8630787},
  doi       = {10.1109/WIFS.2018.8630787},
  bibsource = {dblp computer science bibliography, https://dblp.org}
}

@inproceedings{DBLP:conf/eccv/MasiKMGA20,
  author    = {Iacopo Masi and
               Aditya Killekar and
               Royston Marian Mascarenhas and
               Shenoy Pratik Gurudatt and
               Wael AbdAlmageed},
  editor    = {Andrea Vedaldi and
               Horst Bischof and
               Thomas Brox and
               Jan{-}Michael Frahm},
  title     = {Two-Branch Recurrent Network for Isolating Deepfakes in Videos},
  booktitle = {Computer Vision - {ECCV} 2020 - 16th European Conference, Glasgow,
               UK, August 23-28, 2020, Proceedings, Part {VII}},
  series    = {Lecture Notes in Computer Science},
  volume    = {12352},
  pages     = {667--684},
  publisher = {Springer},
  year      = {2020},
  url       = {https://doi.org/10.1007/978-3-030-58571-6\_39},
  doi       = {10.1007/978-3-030-58571-6\_39},
  bibsource = {dblp computer science bibliography, https://dblp.org}
}

@inproceedings{DBLP:conf/ijcai/ZhangLL0G21,
  author    = {Daichi Zhang and
               Chenyu Li and
               Fanzhao Lin and
               Dan Zeng and
               Shiming Ge},
  editor    = {Zhi{-}Hua Zhou},
  title     = {Detecting Deepfake Videos with Temporal Dropout 3DCNN},
  booktitle = {Proceedings of the Thirtieth International Joint Conference on Artificial
               Intelligence, {IJCAI} 2021, Virtual Event / Montreal, Canada, 19-27
               August 2021},
  pages     = {1288--1294},
  publisher = {ijcai.org},
  year      = {2021},
  url       = {https://doi.org/10.24963/ijcai.2021/178},
  doi       = {10.24963/ijcai.2021/178},
  bibsource = {dblp computer science bibliography, https://dblp.org}
}

@inproceedings{DBLP:conf/icassp/ZhaoYNZ22,
  author    = {Xiaohui Zhao and
               Yang Yu and
               Rongrong Ni and
               Yao Zhao},
  title     = {Exploring Complementarity of Global and Local Spatiotemporal Information
               for Fake Face Video Detection},
  booktitle = {{IEEE} International Conference on Acoustics, Speech and Signal Processing,
               {ICASSP} 2022, Virtual and Singapore, 23-27 May 2022},
  pages     = {2884--2888},
  publisher = {{IEEE}},
  year      = {2022},
  url       = {https://doi.org/10.1109/ICASSP43922.2022.9746061},
  doi       = {10.1109/ICASSP43922.2022.9746061},
  bibsource = {dblp computer science bibliography, https://dblp.org}
}

@inproceedings{DBLP:conf/ijcai/HuXWLW021,
  author    = {Ziheng Hu and
               Hongtao Xie and
               Yuxin Wang and
               Jiahong Li and
               Zhongyuan Wang and
               Yongdong Zhang},
  editor    = {Zhi{-}Hua Zhou},
  title     = {Dynamic Inconsistency-aware DeepFake Video Detection},
  booktitle = {Proceedings of the Thirtieth International Joint Conference on Artificial
               Intelligence, {IJCAI} 2021, Virtual Event / Montreal, Canada, 19-27
               August 2021},
  pages     = {736--742},
  publisher = {ijcai.org},
  year      = {2021},
  url       = {https://doi.org/10.24963/ijcai.2021/102},
  doi       = {10.24963/ijcai.2021/102},
  bibsource = {dblp computer science bibliography, https://dblp.org}
}

@inproceedings{DBLP:conf/aaai/GuCYDLM22,
  author    = {Zhihao Gu and
               Yang Chen and
               Taiping Yao and
               Shouhong Ding and
               Jilin Li and
               Lizhuang Ma},
  title     = {Delving into the Local: Dynamic Inconsistency Learning for DeepFake
               Video Detection},
  booktitle = {Thirty-Sixth {AAAI} Conference on Artificial Intelligence, {AAAI}
               2022, Thirty-Fourth Conference on Innovative Applications of Artificial
               Intelligence, {IAAI} 2022, The Twelveth Symposium on Educational Advances
               in Artificial Intelligence, {EAAI} 2022 Virtual Event, February 22
               - March 1, 2022},
  pages     = {744--752},
  publisher = {{AAAI} Press},
  year      = {2022},
  url       = {https://ojs.aaai.org/index.php/AAAI/article/view/19955},
  bibsource = {dblp computer science bibliography, https://dblp.org}
}

@inproceedings{DBLP:conf/cvpr/LiuLZCH0ZY21,
  author    = {Honggu Liu and
               Xiaodan Li and
               Wenbo Zhou and
               Yuefeng Chen and
               Yuan He and
               Hui Xue and
               Weiming Zhang and
               Nenghai Yu},
  title     = {Spatial-Phase Shallow Learning: Rethinking Face Forgery Detection
               in Frequency Domain},
  booktitle = {{IEEE} Conference on Computer Vision and Pattern Recognition, {CVPR}
               2021, virtual, June 19-25, 2021},
  pages     = {772--781},
  publisher = {Computer Vision Foundation / {IEEE}},
  year      = {2021},
  url       = {https://openaccess.thecvf.com/content/CVPR2021/html/Liu\_Spatial-Phase\_Shallow\_Learning\_Rethinking\_Face\_Forgery\_Detection\_in\_Frequency\_Domain\_CVPR\_2021\_paper.html},
  doi       = {10.1109/CVPR46437.2021.00083},
  bibsource = {dblp computer science bibliography, https://dblp.org}
}

@inproceedings{DBLP:conf/eccv/QianYSCS20,
  author    = {Yuyang Qian and
               Guojun Yin and
               Lu Sheng and
               Zixuan Chen and
               Jing Shao},
  editor    = {Andrea Vedaldi and
               Horst Bischof and
               Thomas Brox and
               Jan{-}Michael Frahm},
  title     = {Thinking in Frequency: Face Forgery Detection by Mining Frequency-Aware
               Clues},
  booktitle = {Computer Vision - {ECCV} 2020 - 16th European Conference, Glasgow,
               UK, August 23-28, 2020, Proceedings, Part {XII}},
  series    = {Lecture Notes in Computer Science},
  volume    = {12357},
  pages     = {86--103},
  publisher = {Springer},
  year      = {2020},
  url       = {https://doi.org/10.1007/978-3-030-58610-2\_6},
  doi       = {10.1007/978-3-030-58610-2\_6},
  bibsource = {dblp computer science bibliography, https://dblp.org}
}

@inproceedings{WildDeepfake,
  title={Wilddeepfake: A challenging real-world dataset for deepfake detection},
  author={Zi, Bojia and Chang, Minghao and Chen, Jingjing and Ma, Xingjun and Jiang, Yu-Gang},
  booktitle={Proceedings of the 28th ACM international conference on multimedia},
  pages={2382--2390},
  year={2020}
}

@inproceedings{Xception,
  author       = {Fran{\c{c}}ois Chollet},
  title        = {Xception: Deep Learning with Depthwise Separable Convolutions},
  booktitle    = {2017 {IEEE} Conference on Computer Vision and Pattern Recognition,
                  {CVPR} 2017, Honolulu, HI, USA, July 21-26, 2017},
  pages        = {1800--1807},
  publisher    = {{IEEE} Computer Society},
  year         = {2017},
  url          = {https://doi.org/10.1109/CVPR.2017.195},
  doi          = {10.1109/CVPR.2017.195},
  bibsource    = {dblp computer science bibliography, https://dblp.org}
}

@inproceedings{meso,
  author       = {Darius Afchar and
                  Vincent Nozick and
                  Junichi Yamagishi and
                  Isao Echizen},
  title        = {MesoNet: a Compact Facial Video Forgery Detection Network},
  booktitle    = {2018 {IEEE} International Workshop on Information Forensics and Security,
                  {WIFS} 2018, Hong Kong, China, December 11-13, 2018},
  pages        = {1--7},
  publisher    = {{IEEE}},
  year         = {2018},
  url          = {https://doi.org/10.1109/WIFS.2018.8630761},
  doi          = {10.1109/WIFS.2018.8630761},
  bibsource    = {dblp computer science bibliography, https://dblp.org}
}

@inproceedings{DBLP:conf/cvpr/LiL19c,
  author       = {Yuezun Li and
                  Siwei Lyu},
  title        = {Exposing DeepFake Videos By Detecting Face Warping Artifacts},
  booktitle    = {{IEEE} Conference on Computer Vision and Pattern Recognition Workshops,
                  {CVPR} Workshops 2019, Long Beach, CA, USA, June 16-20, 2019},
  pages        = {46--52},
  publisher    = {Computer Vision Foundation / {IEEE}},
  year         = {2019},
  url          = {http://openaccess.thecvf.com/content\_CVPRW\_2019/html/Media\_Forensics/Li\_Exposing\_DeepFake\_Videos\_By\_Detecting\_Face\_Warping\_Artifacts\_CVPRW\_2019\_paper.html},
  bibsource    = {dblp computer science bibliography, https://dblp.org}
}

@inproceedings{FTCN,
  author       = {Yinglin Zheng and
                  Jianmin Bao and
                  Dong Chen and
                  Ming Zeng and
                  Fang Wen},
  title        = {Exploring Temporal Coherence for More General Video Face Forgery Detection},
  booktitle    = {2021 {IEEE/CVF} International Conference on Computer Vision, {ICCV}
                  2021, Montreal, QC, Canada, October 10-17, 2021},
  pages        = {15024--15034},
  publisher    = {{IEEE}},
  year         = {2021},
  url          = {https://doi.org/10.1109/ICCV48922.2021.01477},
  doi          = {10.1109/ICCV48922.2021.01477},
  bibsource    = {dblp computer science bibliography, https://dblp.org}
}

@inproceedings{ICT,
  author       = {Xiaoyi Dong and
                  Jianmin Bao and
                  Dongdong Chen and
                  Ting Zhang and
                  Weiming Zhang and
                  Nenghai Yu and
                  Dong Chen and
                  Fang Wen and
                  Baining Guo},
  title        = {Protecting Celebrities from DeepFake with Identity Consistency Transformer},
  booktitle    = {{IEEE/CVF} Conference on Computer Vision and Pattern Recognition,
                  {CVPR} 2022, New Orleans, LA, USA, June 18-24, 2022},
  pages        = {9458--9468},
  publisher    = {{IEEE}},
  year         = {2022},
  url          = {https://doi.org/10.1109/CVPR52688.2022.00925},
  doi          = {10.1109/CVPR52688.2022.00925},
  bibsource    = {dblp computer science bibliography, https://dblp.org}
}

@inproceedings{ID-Reveal,
  author       = {Davide Cozzolino and
                  Andreas R{\"{o}}ssler and
                  Justus Thies and
                  Matthias Nie{\ss}ner and
                  Luisa Verdoliva},
  title        = {ID-Reveal: Identity-aware DeepFake Video Detection},
  booktitle    = {2021 {IEEE/CVF} International Conference on Computer Vision, {ICCV}
                  2021, Montreal, QC, Canada, October 10-17, 2021},
  pages        = {15088--15097},
  publisher    = {{IEEE}},
  year         = {2021},
  url          = {https://doi.org/10.1109/ICCV48922.2021.01483},
  doi          = {10.1109/ICCV48922.2021.01483},
  bibsource    = {dblp computer science bibliography, https://dblp.org}
}

@article{OuterFace,
  author       = {Xiaoyi Dong and
                  Jianmin Bao and
                  Dongdong Chen and
                  Weiming Zhang and
                  Nenghai Yu and
                  Dong Chen and
                  Fang Wen and
                  Baining Guo},
  title        = {Identity-Driven DeepFake Detection},
  journal      = {CoRR},
  volume       = {abs/2012.03930},
  year         = {2020},
  url          = {https://arxiv.org/abs/2012.03930},
  eprinttype    = {arXiv},
  eprint       = {2012.03930},
  bibsource    = {dblp computer science bibliography, https://dblp.org}
}

@inproceedings{DBLP:conf/cvpr/HuangWYA00Y23,
  author       = {Baojin Huang and
                  Zhongyuan Wang and
                  Jifan Yang and
                  Jiaxin Ai and
                  Qin Zou and
                  Qian Wang and
                  Dengpan Ye},
  title        = {Implicit Identity Driven Deepfake Face Swapping Detection},
  booktitle    = {{IEEE/CVF} Conference on Computer Vision and Pattern Recognition,
                  {CVPR} 2023, Vancouver, BC, Canada, June 17-24, 2023},
  pages        = {4490--4499},
  publisher    = {{IEEE}},
  year         = {2023},
  url          = {https://doi.org/10.1109/CVPR52729.2023.00436},
  doi          = {10.1109/CVPR52729.2023.00436},
  bibsource    = {dblp computer science bibliography, https://dblp.org}
}

@inproceedings{ucf,
  author       = {Zhiyuan Yan and
                  Yong Zhang and
                  Yanbo Fan and
                  Baoyuan Wu},
  title        = {{UCF:} Uncovering Common Features for Generalizable Deepfake Detection},
  booktitle    = {{IEEE/CVF} International Conference on Computer Vision, {ICCV} 2023,
                  Paris, France, October 1-6, 2023},
  pages        = {22355--22366},
  publisher    = {{IEEE}},
  year         = {2023},
  url          = {https://doi.org/10.1109/ICCV51070.2023.02048},
  doi          = {10.1109/ICCV51070.2023.02048},
  bibsource    = {dblp computer science bibliography, https://dblp.org}
}

@article{foteinopoulou2024hitchhiker,
  title={A Hitchhiker's Guide to Fine-Grained Face Forgery Detection Using Common Sense Reasoning},
  author={Foteinopoulou, Niki M and Ghorbel, Enjie and Aouada, Djamila},
  journal={Advances in Neural Information Processing Systems},
  volume={37},
  pages={2943--2976},
  year={2024}
}

@inproceedings{DBLP:conf/eccv/ZhangCGSB24,
  author       = {Yue Zhang and
                  Ben Colman and
                  Xiao Guo and
                  Ali Shahriyari and
                  Gaurav Bharaj},
  editor       = {Ales Leonardis and
                  Elisa Ricci and
                  Stefan Roth and
                  Olga Russakovsky and
                  Torsten Sattler and
                  G{\"{u}}l Varol},
  title        = {Common Sense Reasoning for Deepfake Detection},
  booktitle    = {Computer Vision - {ECCV} 2024 - 18th European Conference, Milan, Italy,
                  September 29-October 4, 2024, Proceedings, Part {LXXXVIII}},
  series       = {Lecture Notes in Computer Science},
  volume       = {15146},
  pages        = {399--415},
  publisher    = {Springer},
  year         = {2024},
  url          = {https://doi.org/10.1007/978-3-031-73223-2\_22},
  doi          = {10.1007/978-3-031-73223-2\_22},
  bibsource    = {dblp computer science bibliography, https://dblp.org}
}

@article{ffaa,
  title={Ffaa: Multimodal large language model based explainable open-world face forgery analysis assistant},
  author={Huang, Zhengchao and Xia, Bin and Lin, Zicheng and Mou, Zhun and Yang, Wenming and Jia, Jiaya},
  journal={arXiv preprint arXiv:2408.10072},
  year={2024}
}

@article{fakeshield,
  title={Fakeshield: Explainable image forgery detection and localization via multi-modal large language models},
  author={Xu, Zhipei and Zhang, Xuanyu and Li, Runyi and Tang, Zecheng and Huang, Qing and Zhang, Jian},
  journal={ICLR},
  year={2025}
}

@inproceedings{Can_ChatGPT_Detect,
  author       = {Shan Jia and
                  Reilin Lyu and
                  Kangran Zhao and
                  Yize Chen and
                  Zhiyuan Yan and
                  Yan Ju and
                  Chuanbo Hu and
                  Xin Li and
                  Baoyuan Wu and
                  Siwei Lyu},
  title        = {Can ChatGPT Detect DeepFakes? {A} Study of Using Multimodal Large
                  Language Models for Media Forensics},
  booktitle    = {{IEEE/CVF} Conference on Computer Vision and Pattern Recognition,
                  {CVPR} 2024 - Workshops, Seattle, WA, USA, June 17-18, 2024},
  pages        = {4324--4333},
  publisher    = {{IEEE}},
  year         = {2024},
  url          = {https://doi.org/10.1109/CVPRW63382.2024.00436},
  doi          = {10.1109/CVPRW63382.2024.00436},
  bibsource    = {dblp computer science bibliography, https://dblp.org}
}

@inproceedings{IDForge,
  author       = {Junhao Xu and
                  Jingjing Chen and
                  Xue Song and
                  Feng Han and
                  Haijun Shan and
                  Yu{-}Gang Jiang},
  editor       = {Jianfei Cai and
                  Mohan S. Kankanhalli and
                  Balakrishnan Prabhakaran and
                  Susanne Boll and
                  Ramanathan Subramanian and
                  Liang Zheng and
                  Vivek K. Singh and
                  Pablo C{\'{e}}sar and
                  Lexing Xie and
                  Dong Xu},
  title        = {Identity-Driven Multimedia Forgery Detection via Reference Assistance},
  booktitle    = {Proceedings of the 32nd {ACM} International Conference on Multimedia,
                  {MM} 2024, Melbourne, VIC, Australia, 28 October 2024 - 1 November
                  2024},
  pages        = {3887--3896},
  publisher    = {{ACM}},
  year         = {2024},
  url          = {https://doi.org/10.1145/3664647.3680622},
  doi          = {10.1145/3664647.3680622},
  bibsource    = {dblp computer science bibliography, https://dblp.org}
}

@inproceedings{recce,
  author       = {Junyi Cao and
                  Chao Ma and
                  Taiping Yao and
                  Shen Chen and
                  Shouhong Ding and
                  Xiaokang Yang},
  title        = {End-to-End Reconstruction-Classification Learning for Face Forgery Detection},
  booktitle    = {{IEEE/CVF} Conference on Computer Vision and Pattern Recognition,
                  {CVPR} 2022, New Orleans, LA, USA, June 18-24, 2022},
  pages        = {4103--4112},
  publisher    = {{IEEE}},
  year         = {2022},
  url          = {https://doi.org/10.1109/CVPR52688.2022.00408},
  doi          = {10.1109/CVPR52688.2022.00408},
  bibsource    = {dblp computer science bibliography, https://dblp.org}
}

@article{gpt4o,
  author       = {Aaron Hurst and
                  Adam Lerer and
                  Adam P. Goucher and
                  Adam Perelman and
                  Aditya Ramesh and
                  Aidan Clark and
                  AJ Ostrow and
                  Akila Welihinda and
                  Alan Hayes and
                  Alec Radford and
                  Aleksander Madry and
                  Alex Baker{-}Whitcomb and
                  Alex Beutel and
                  Alex Borzunov and
                  Alex Carney and
                  Alex Chow and
                  Alex Kirillov and
                  Alex Nichol and
                  Alex Paino and
                  Alex Renzin and
                  Alex Tachard Passos and
                  Alexander Kirillov and
                  Alexi Christakis and
                  Alexis Conneau and
                  Ali Kamali and
                  Allan Jabri and
                  Allison Moyer and
                  Allison Tam and
                  Amadou Crookes and
                  Amin Tootoonchian and
                  Ananya Kumar and
                  Andrea Vallone and
                  Andrej Karpathy and
                  Andrew Braunstein and
                  Andrew Cann and
                  Andrew Codispoti and
                  Andrew Galu and
                  Andrew Kondrich and
                  Andrew Tulloch and
                  Andrey Mishchenko and
                  Angela Baek and
                  Angela Jiang and
                  Antoine Pelisse and
                  Antonia Woodford and
                  Anuj Gosalia and
                  Arka Dhar and
                  Ashley Pantuliano and
                  Avi Nayak and
                  Avital Oliver and
                  Barret Zoph and
                  Behrooz Ghorbani and
                  Ben Leimberger and
                  Ben Rossen and
                  Ben Sokolowsky and
                  Ben Wang and
                  Benjamin Zweig and
                  Beth Hoover and
                  Blake Samic and
                  Bob McGrew and
                  Bobby Spero and
                  Bogo Giertler and
                  Bowen Cheng and
                  Brad Lightcap and
                  Brandon Walkin and
                  Brendan Quinn and
                  Brian Guarraci and
                  Brian Hsu and
                  Bright Kellogg and
                  Brydon Eastman and
                  Camillo Lugaresi and
                  Carroll L. Wainwright and
                  Cary Bassin and
                  Cary Hudson and
                  Casey Chu and
                  Chad Nelson and
                  Chak Li and
                  Chan Jun Shern and
                  Channing Conger and
                  Charlotte Barette and
                  Chelsea Voss and
                  Chen Ding and
                  Cheng Lu and
                  Chong Zhang and
                  Chris Beaumont and
                  Chris Hallacy and
                  Chris Koch and
                  Christian Gibson and
                  Christina Kim and
                  Christine Choi and
                  Christine McLeavey and
                  Christopher Hesse and
                  Claudia Fischer and
                  Clemens Winter and
                  Coley Czarnecki and
                  Colin Jarvis and
                  Colin Wei and
                  Constantin Koumouzelis and
                  Dane Sherburn},
  title        = {GPT-4o System Card},
  journal      = {CoRR},
  volume       = {abs/2410.21276},
  year         = {2024},
  url          = {https://doi.org/10.48550/arXiv.2410.21276},
  doi          = {10.48550/ARXIV.2410.21276},
  eprinttype    = {arXiv},
  eprint       = {2410.21276},
  bibsource    = {dblp computer science bibliography, https://dblp.org}
}

@article{qwen25vl,
  author       = {Shuai Bai and
                  Keqin Chen and
                  Xuejing Liu and
                  Jialin Wang and
                  Wenbin Ge and
                  Sibo Song and
                  Kai Dang and
                  Peng Wang and
                  Shijie Wang and
                  Jun Tang and
                  Humen Zhong and
                  Yuanzhi Zhu and
                  Ming{-}Hsuan Yang and
                  Zhaohai Li and
                  Jianqiang Wan and
                  Pengfei Wang and
                  Wei Ding and
                  Zheren Fu and
                  Yiheng Xu and
                  Jiabo Ye and
                  Xi Zhang and
                  Tianbao Xie and
                  Zesen Cheng and
                  Hang Zhang and
                  Zhibo Yang and
                  Haiyang Xu and
                  Junyang Lin},
  title        = {Qwen2.5-VL Technical Report},
  journal      = {CoRR},
  volume       = {abs/2502.13923},
  year         = {2025},
  url          = {https://doi.org/10.48550/arXiv.2502.13923},
  doi          = {10.48550/ARXIV.2502.13923},
  eprinttype    = {arXiv},
  eprint       = {2502.13923},
  bibsource    = {dblp computer science bibliography, https://dblp.org}
}

@inproceedings{llava,
  author       = {Haotian Liu and
                  Chunyuan Li and
                  Qingyang Wu and
                  Yong Jae Lee},
  editor       = {Alice Oh and
                  Tristan Naumann and
                  Amir Globerson and
                  Kate Saenko and
                  Moritz Hardt and
                  Sergey Levine},
  title        = {Visual Instruction Tuning},
  booktitle    = {Advances in Neural Information Processing Systems 36: Annual Conference
                  on Neural Information Processing Systems 2023, NeurIPS 2023, New Orleans,
                  LA, USA, December 10 - 16, 2023},
  year         = {2023},
  url          = {http://papers.nips.cc/paper\_files/paper/2023/hash/6dcf277ea32ce3288914faf369fe6de0-Abstract-Conference.html},
  bibsource    = {dblp computer science bibliography, https://dblp.org}
}

@inproceedings{yollava,
  author       = {Thao Nguyen and
                  Haotian Liu and
                  Yuheng Li and
                  Mu Cai and
                  Utkarsh Ojha and
                  Yong Jae Lee},
  editor       = {Amir Globersons and
                  Lester Mackey and
                  Danielle Belgrave and
                  Angela Fan and
                  Ulrich Paquet and
                  Jakub M. Tomczak and
                  Cheng Zhang},
  title        = {Yo'LLaVA: Your Personalized Language and Vision Assistant},
  booktitle    = {Advances in Neural Information Processing Systems 38: Annual Conference
                  on Neural Information Processing Systems 2024, NeurIPS 2024, Vancouver,
                  BC, Canada, December 10 - 15, 2024},
  year         = {2024},
  url          = {http://papers.nips.cc/paper\_files/paper/2024/hash/48088756ec0ce6ba362bddc7ebeb3915-Abstract-Conference.html},
  bibsource    = {dblp computer science bibliography, https://dblp.org}
}

@article{ESSP,
  author       = {Jiaxuan Chen and
                  Jieteng Yao and
                  Li Niu},
  title        = {A Single Simple Patch is All You Need for AI-generated Image Detection},
  journal      = {CoRR},
  volume       = {abs/2402.01123},
  year         = {2024},
  url          = {https://doi.org/10.48550/arXiv.2402.01123},
  doi          = {10.48550/ARXIV.2402.01123},
  eprinttype    = {arXiv},
  eprint       = {2402.01123},
  bibsource    = {dblp computer science bibliography, https://dblp.org}
}

@inproceedings{NPR,
  author       = {Chuangchuang Tan and
                  Huan Liu and
                  Yao Zhao and
                  Shikui Wei and
                  Guanghua Gu and
                  Ping Liu and
                  Yunchao Wei},
  title        = {Rethinking the Up-Sampling Operations in CNN-Based Generative Network
                  for Generalizable Deepfake Detection},
  booktitle    = {{IEEE/CVF} Conference on Computer Vision and Pattern Recognition,
                  {CVPR} 2024, Seattle, WA, USA, June 16-22, 2024},
  pages        = {28130--28139},
  publisher    = {{IEEE}},
  year         = {2024},
  url          = {https://doi.org/10.1109/CVPR52733.2024.02657},
  doi          = {10.1109/CVPR52733.2024.02657},
  bibsource    = {dblp computer science bibliography, https://dblp.org}
}

@inproceedings{photomaker,
  author       = {Zhen Li and
                  Mingdeng Cao and
                  Xintao Wang and
                  Zhongang Qi and
                  Ming{-}Ming Cheng and
                  Ying Shan},
  title        = {PhotoMaker: Customizing Realistic Human Photos via Stacked {ID} Embedding},
  booktitle    = {{IEEE/CVF} Conference on Computer Vision and Pattern Recognition,
                  {CVPR} 2024, Seattle, WA, USA, June 16-22, 2024},
  pages        = {8640--8650},
  publisher    = {{IEEE}},
  year         = {2024},
  url          = {https://doi.org/10.1109/CVPR52733.2024.00825},
  doi          = {10.1109/CVPR52733.2024.00825},
  bibsource    = {dblp computer science bibliography, https://dblp.org}
}

@article{storymaker,
  author       = {Zhengguang Zhou and
                  Jing Li and
                  Huaxia Li and
                  Nemo Chen and
                  Xu Tang},
  title        = {StoryMaker: Towards Holistic Consistent Characters in Text-to-image
                  Generation},
  journal      = {CoRR},
  volume       = {abs/2409.12576},
  year         = {2024},
  url          = {https://doi.org/10.48550/arXiv.2409.12576},
  doi          = {10.48550/ARXIV.2409.12576},
  eprinttype    = {arXiv},
  eprint       = {2409.12576},
  bibsource    = {dblp computer science bibliography, https://dblp.org}
}

@inproceedings{pulid,
  author       = {Zinan Guo and
                  Yanze Wu and
                  Zhuowei Chen and
                  Lang Chen and
                  Peng Zhang and
                  Qian He},
  editor       = {Amir Globersons and
                  Lester Mackey and
                  Danielle Belgrave and
                  Angela Fan and
                  Ulrich Paquet and
                  Jakub M. Tomczak and
                  Cheng Zhang},
  title        = {PuLID: Pure and Lightning {ID} Customization via Contrastive Alignment},
  booktitle    = {Advances in Neural Information Processing Systems 38: Annual Conference
                  on Neural Information Processing Systems 2024, NeurIPS 2024, Vancouver,
                  BC, Canada, December 10 - 15, 2024},
  year         = {2024},
  url          = {http://papers.nips.cc/paper\_files/paper/2024/hash/409fcc9d24b549969b8b9be68b56a7be-Abstract-Conference.html},
  bibsource    = {dblp computer science bibliography, https://dblp.org}
}

@inproceedings{myvlm,
  author       = {Yuval Alaluf and
                  Elad Richardson and
                  Sergey Tulyakov and
                  Kfir Aberman and
                  Daniel Cohen{-}Or},
  editor       = {Ales Leonardis and
                  Elisa Ricci and
                  Stefan Roth and
                  Olga Russakovsky and
                  Torsten Sattler and
                  G{\"{u}}l Varol},
  title        = {MyVLM: Personalizing VLMs for User-Specific Queries},
  booktitle    = {Computer Vision - {ECCV} 2024 - 18th European Conference, Milan, Italy,
                  September 29-October 4, 2024, Proceedings, Part {XIII}},
  series       = {Lecture Notes in Computer Science},
  volume       = {15071},
  pages        = {73--91},
  publisher    = {Springer},
  year         = {2024},
  url          = {https://doi.org/10.1007/978-3-031-72624-8\_5},
  doi          = {10.1007/978-3-031-72624-8\_5},
  bibsource    = {dblp computer science bibliography, https://dblp.org}
}

@article{lumen,
  title={Lumen: Unleashing versatile vision-centric capabilities of large multimodal models},
  author={Jiao, Yang and Chen, Shaoxiang and Jie, Zequn and Chen, Jingjing and Ma, Lin and Jiang, Yu-Gang},
  journal={arXiv preprint arXiv:2403.07304},
  year={2024}
}

@article{unitoken,
  title={UniToken: Harmonizing Multimodal Understanding and Generation through Unified Visual Encoding},
  author={Jiao, Yang and Qiu, Haibo and Jie, Zequn and Chen, Shaoxiang and Chen, Jingjing and Ma, Lin and Jiang, Yu-Gang},
  journal={arXiv preprint arXiv:2504.04423},
  year={2025}
}

@article{eagle,
  title={Eagle: Towards efficient arbitrary referring visual prompts comprehension for multimodal large language models},
  author={Zhang, Jiacheng and Jiao, Yang and Chen, Shaoxiang and Chen, Jingjing and Jiang, Yu-Gang},
  journal={arXiv preprint arXiv:2409.16723},
  year={2024}
}

\end{document}